\documentclass[12pt,a4paper]{revtex4}
\usepackage{amsmath}  
\usepackage{amssymb}  
\usepackage{dsfont}   

\begin{document}
\title{The position-momentum symmetry principle}
\author{Alberto C. de la Torre}\email{delatorre@mdp.edu.ar}
\affiliation{ IFIMAR - (CONICET-UNMDP) \\ Departamento de F\'{\i}sica -
 Facultad de Ciencias Exactas y Naturales -
 Universidad Nacional de Mar del Plata\\
 Funes 3350, 7600 Mar del Plata, Argentina.}
\begin{abstract}It is shown that the Fourier transformation that
relates position and momentum representations of quantum mechanics
can be understood as a consequence of a symmetry principle that
establishes the equivalence of being and becoming in the description
of reality. There are however other transformation compatible with
the same principle that could lead to different formalisms of
quantum mechanics.
\\ Published in nnnn \textbf{nn},  nn–nn, (2013). arXiv: nnnnn.nnnnn
\end{abstract}
\maketitle
\section{introduction}
In the attempt to develop an understanding of quantum mechanics it
is convenient to treat the simplest possible quantum system
consisting in a structureless free particle moving in one dimension.
This system has only two relevant observables: position and
momentum. Their relation in classical mechanics has several degrees
of abstraction: from the simplest definition of momentum as $p=mv$
to the relevant conjugate variables in the Legendre transformation
of the Lagrangian in order to obtain the Hamiltonian. In any case,
their conceptual meaning is related to the space-time location of
matter and its movement or evolution: being and becoming.

In quantum mechanics these observables are less well understood. Not
only they have the usual indetermination, characteristic of all
quantum observables, but in addition they acquire a mysterious
correlation that is essentially formalized in Heisenberg's
uncertainty principle: we have gotten used to it, but we don't
really understand it.

There are several features of the formalism of quantum mechanics
concerning the observables of position and momentum that are closely
related and are just different manifestations of some profound
relation between these two fundamental observables. These
are:\cite{dlt0}
\begin{itemize}
\item The commutator between position and momentum is $[X,P]=i$ (we
assume units with $\hbar=1$).
\item The momentum operator is the generator of translations because, for
any position dependent operator $F(X)$, we have
$[F,P]=i\frac{dF}{dX}$ and therefore $U^{\dag}_{a}F(X)U_{a}=F(X+a)$,
where $U_{a}=\exp(-iaP)$.  In a similar way, position is the
generator of impulsions.
\item The internal product between the basis elements corresponding to
position and momentum operators, $\{\varphi_{x}\}$ and
$\{\phi_{p}\}$, is
$\langle\varphi_{x},\phi_{p}\rangle=\frac{1}{\sqrt{2\pi}}\exp(ixp)$.
This implies that position and momentum representations are related
by the Fourier transformation. The fact that the bases  are
\emph{unbiased}, that is, $|\langle\varphi_{x},\phi_{p}\rangle|$ is
independent of $x$ and $p$, is a manifestation of the physical
independence of position and momentum observables.
\end{itemize}
The three statements above are equivalent since from any of them one
can prove the others and they contain all the essential relations
between position and momentum. Therefore if we manage to develop an
understanding or an interpretation of any of them we gain a deep
insight in this central point of quantum mechanics. In this note we
attempt to give a physical meaning to the third feature: we will
find out why position and momentum representations are related by
the Fourier transformation. For this we postulate a symmetry
principle that establishes the equivalence of the being and becoming
of a physical system. In particular, we propose that the roles of
position and momentum can be interchanged or permuted.

\section{PHILOSOPHICAL ROOTS}
Being and becoming were the main issues in the opposing ontologies
of Heraclitus and Parmenides in the fifth century before the
Christian Era. For Heraclitus, nothing in the world is constant;
\emph{``everything flows''} and the unique reality is the permanent
change. Parmenides, on the contrary, claimed that all change is
impossible because \emph{nothing can stop being what it is in order
to start being what is not}. For him all changes are a dilution of
our senses: existence is timeless and unchanging. Accordingly, his
disciple Zeno devised paradoxes tending to show that all motion is
impossible.  A somewhat conciliating view was adopted by Leucippus
(that with his pupil Democritus developed the atomistic hypothesis)
assuming that atoms are permanent but can change in their movement
and combinations.

Apparently, in the antique Greek philosophy it was important to
establish some sort of priority between the different views of
reality as permanent and unchanging or as in a continuous change.
For Heraclitus becoming has priority, Parmenides denies every change
and Leucippus favours the being of atoms but does not deny their
change. Today, the question of priority between being and becoming
is not a problem that requires a solution: reality may well be a non
contradicting, complementary, combination of being and becoming from
which we perceive different perspectives, sometimes exhibiting the
being and other times the becoming. Each perspective brings a
complete, but not unique, view of reality and both perspectives are
equivalent in the sense that from the being we can derive its
becoming and \textit{vice-versa}. This is precisely what we do in
elementary calculus: from a function we can derive its changes
(derivatives) and from the changes we can obtain the function
(integration).

In this work we postulate a symmetry principle based in the
equivalence of being and becoming: these are two different, but
complementary, perspectives to approach reality. In the case of
classical mechanics, this principle is formalized in a well known
canonical transformation and therefore it does not bring anything
really new, but in quantum mechanics it provides a novel
understanding or explanation for the relation between position and
momentum that is at the basis of all essential features of the
theory.
\section{Being and becoming in mechanics}
We can identify the two different philosophical considerations of
reality --being and becoming-- with two different perspectives in
the analysis of a physical system: the space-time or the
energy-momentum point of view. The being-becoming symmetry principle
means for mechanics the equivalence of description through
space-time or energy-momentum.

For the case of the simplest physical system consisting in a moving
structureless particle in one dimension, the description of the
system by means of its time depending coordinate $x$ is equivalent
to the corresponding description by means of its momentum $p$. The
symmetry principle, in this simplest case, establishes then that the
roles of position and momentum can be interchanged.

In classical mechanics the symmetry principle is not very
surprising. It is, in fact, well known that there is a
transformation $x\rightarrow x'= p$ and $p\rightarrow p'= -x$ that
leaves the Poisson brackets invariant. This canonical transformation
is a special case of a family of transformations that amount to a
rotation of the phase space by an arbitrary angle $x\rightarrow x'=
x \cos\theta + p\sin\theta$ and $p\rightarrow p'= -x\sin\theta +
p\cos\theta$. For $\theta=\frac{\pi}{2}$ we obtain the
transformation mentioned before and for  $\theta=\pi$ we get the
parity transformation $x\rightarrow -x, \ p\rightarrow -p$.

We will now see that in quantum mechanics the symmetry principle has
an interesting consequence. The quantum mechanical predictions for
position and momentum are given by the distribution of their
eigenvalues  $\rho (x)$ and $\varpi (p)$. These are non-negative and
normalized, in the sense that their integration is unity. They are
measured in an experiment by the frequency of appearance of each
eigenvalue, however, in all rigour, they are not probability
distributions and a proper name for them could be the
\emph{existential weight} for position and momentum\cite{dlt1}.
Unfortunately, the misnomer \emph{probability distribution} is
irreversibly installed in quantum mechanics.

One striking feature of quantum mechanics is that the complete
knowledge of both existential weights $\rho (x)$ and $\varpi (p)$ is
\emph{not} sufficient for an unambiguous determination of the state
of the system, although position and momentum are the unique
relevant observables. Therefore these distributions do not provide a
complete description of the system. This fact, known as the
\emph{Pauli problem}\cite{pau}, has triggered much activity in
trying to establish sufficient conditions for state determination; a
problem still
unsolved\cite{wei1,wei2,qinf,unbiaBas1,unbiaBas2,dlt4}. In order to
fix the state of the system we need, besides the distributions $\rho
(x)$ or $\varpi (p)$, something more.

From the position point of view, besides the information contained
in the existential weight $\rho (x)$ we define another function
$\alpha(x)$ that somehow encodes all the missing information about
all other observables of the system (functions of position and
momentum). Furthermore, we can combine both functions in a single
complex function $f(x)$ that is a candidate for the state of the
system because it contains information on all relevant observables.
A representation for the state of the system is then
\begin{equation}\label{stateposition}
f(x) =\sqrt{\rho (x)}\exp \left(i\alpha(x)\right)\ .
\end{equation}
In the same way, but from the momentum perspective, we obtain
another representation for the state
\begin{equation}\label{statemomentum}
g(p) =\sqrt{\varpi (p) }\exp\left( i\beta(p)\right)\ .
\end{equation}
The squared roots in these definitions were introduced because it is
convenient to consider these two normalized complex functions as
members of the Hilbert space $\mathfrak{L}_{2}$ with unit norm:
$\|f\|^{2}=\langle f,f\rangle = 1$  and $\|g\|^{2}=\langle
g,g\rangle = 1$.

Both functions contain all relevant information about the system and
therefore they are redundant. There must exist then an
\emph{invertible operator} $\mathfrak{F}$ that relates them:
\begin{equation}\label{fourier}
 g = \mathfrak{F} f \ .
\end{equation}
We will now prove that the being-becoming symmetry principle implies
some conditions on $\mathfrak{F}$ that are satisfied by the Fourier
transformation.

Let us assume then a state of the system given, redundantly, by
$f(x)$ and $g(p)$. The symmetry principle under the transformation
$x\rightarrow x'=p$ and $p \rightarrow p'=-x$ means that another
possible state of the system is given by $f'(x) = g(x)$ and
$g'(p)=f(-p)=\mathcal{P}f(p)$, where we have applied the
\emph{parity operator} $\mathcal{P}$ to perform the change of sign.
As a consequence of this symmetry principle, $f$ and $g$ as well as
$f'$ and $g'$ are related by the same operator $\mathfrak{F}$.
Therefore we have $g'= \mathfrak{F} f' = \mathfrak{F} g=
\mathfrak{F}\mathfrak{F}f$. Replacing $g'=\mathcal{P}f$ and
considering that $f$ is arbitrary we conclude that
\begin{equation}\label{Fou2eqPar}
\mathfrak{F}^{2}=\mathcal{P}\ .
\end{equation}
Now, since $\mathcal{P}$ is an involution,
$\mathcal{P}^{2}=\mathds{1}$, the transformation $\mathfrak{F}$ has
period 4:
\begin{equation}\label{per4}
 \mathfrak{F}^{4}=\mathds{1}\ ,
\end{equation}
its eigenvalues are $\{1,-i,-1,i\}$ and the inverse is
$\mathfrak{F}^{-1}=\mathfrak{F}^{3}$.

It is easy to check that the Fourier transformation
\begin{equation}\label{Fourtransf}
 (\mathfrak{F}f)(p) =\frac{1}{\sqrt{2\pi}} \int^{\infty}_{-\infty} dx\ \exp(-ipx) f(x)\
\end{equation}
satisfies all these requirements and therefore we can take the
Fourier transformation for the operator $\mathfrak{F}$ that relates
the position and  momentum perspectives of the physical system. It
is interesting to notice however that the Fourier transformation is
not the unique transformation that satisfies the requirements of the
proposed symmetry principle. In the appendix it is shown that there
are in fact infinite many transformations $\mathfrak{K}$,  different
from the Fourier transformation $\mathfrak{F}$, having the same
properties $\mathfrak{K}^{2}=\mathcal{P}$ and
$\mathfrak{K}^{4}=\mathds{1}$.
\section{FINAL COMMENTS}
We have seen that the symmetry principle that establishes an
equivalence of position and momentum imposes some requirements on
the formalism that are satisfied by the Fourier transformation
between the position and momentum pictures or perspectives of the
quantum system. In this sense, the Fourier transformation in quantum
mechanics is related to the philosophical equivalence of being and
becoming.

However, we have also seen that the Fourier transformation is not
the unique one to have these properties and a natural question is if
one can develop a different formalism of quantum mechanics based on
some other transformation different from the Fourier one. In this
case many questions arise: what predictions do these other theories
make? are they the same as the conventional quantum mechanics? do
all these other theories have some form of an uncertainty principle?
what are the commutation relations of position and momentum for
these other theories? etc.

In a different approach, one can try to find an additional physical
principle that excludes all other transformation leaving the Fourier
transformation as the unique one acceptable for quantum mechanics.

I would like to thank M. C. Moure for her comments and suggestions.

\section{appendix}
We prove here that there are infinite many transformations
$\mathfrak{K}$, different from the Fourier transformation
$\mathfrak{F}$, with the property $\mathfrak{K}^{2}=\mathcal{P}$ and
$\mathfrak{K}^{4}=\mathds{1}$.

Let $\{\psi_{n}\}\ n=0,1,2,\ldots$ be the orthonormal basis in
$\mathfrak{L}_{2}$ built with the Hermite Functions
\begin{equation}\label{hermite}
 \psi_{n}(x)=\exp(-\frac{x^{2}}{2})H_{n}(x)\ ,\
 H_{n}(x)=(-1)^{n}\exp(x^{2})\left(\frac{d}{dx}\right)^{n}\exp(-x^{2})\
 .
 \end{equation}
These are the eigenvectors of the Fourier transformation operator:
\begin{equation}\label{fourier}
 \mathfrak{F}\psi_{n}=(-i)^{n}\psi_{n}\ .
\end{equation}
Let us consider a decomposition of $\mathfrak{L}_{2}$ in the
invariant subspaces corresponding to the four degenerate eigenvalues
$\{1,-i,-1,i\}$,
$\mathfrak{L}_{2}=\mathcal{H}_{0}\oplus\mathcal{H}_{1}
\oplus\mathcal{H}_{2}\oplus\mathcal{H}_{3}$ where the subspaces
$\mathcal{H}_{k}\ k=0,1,2,3$ are spanned by the subbases
$\{\psi_{4r+k}\}\ r=0,1,2,\ldots$ Notice that all functions in
$\mathcal{H}_{0}$ and $\mathcal{H}_{2}$ are \emph{even} whereas
those in $\mathcal{H}_{1}$ and $\mathcal{H}_{3}$ are \emph{odd}
under the parity transformation $\mathcal{P}$. Let $P_{k} $ be the
projectors in the subspaces $\mathcal{H}_{k}$. With this, the
Fourier operator has a spectral decomposition $\mathfrak{F}=
P_{0}-iP_{1}-P_{2}+iP_{3}$ and one can easily check that it has the
properties $\mathfrak{F}^{2}=\mathcal{P}$ and
$\mathfrak{F}^{4}=\mathds{1}$.

Now let us build another decomposition of the \emph{even} subspace
$\mathcal{H}_{0}\oplus\mathcal{H}_{2}$ by choosing an arrangement of
the basis elements different from $\{\psi_{4r}\}\ r=0,1,2,\ldots$ of
$\mathcal{H}_{0}$ and $\{\psi_{4r+2}\}\ r=0,1,2,\ldots$ of
$\mathcal{H}_{2}$. For instance, we can take randomly two disjoint
sets of indices from $\{0,2,4,6,8,10,12\ldots\}$ in order to build
the two sub bases for the decomposition
$\mathcal{\widetilde{H}}_{0}\oplus\mathcal{\widetilde{H}}_{2}$.
There are infinite ways to chose the decomposition
$\mathcal{\widetilde{H}}_{0}\oplus\mathcal{\widetilde{H}}_{2}$ and
similarly we make a different decomposition for the \emph{odd} space
$\mathcal{H}_{1}\oplus\mathcal{H}_{3}$ resulting in
$\mathcal{\widetilde{H}}_{1}\oplus\mathcal{\widetilde{H}}_{3}$.

Let $\widetilde{P}_{k} $ be the projectors in the subspaces
$\mathcal{\widetilde{H}}_{k}$. With this, we define an operator
$\mathfrak{K}=
\widetilde{P}_{0}-i\widetilde{P}_{1}-\widetilde{P}_{2}+i\widetilde{P}_{3}$,
that is clearly different from $\mathfrak{F}$ because it has
different invariant subspaces, and one can easily check that it has
the properties $\mathfrak{K}^{2}=\mathcal{P}$ and
$\mathfrak{K}^{4}=\mathds{1}$.


\begin{thebibliography}{99}
\bibitem{dlt0}
A. C. de la Torre, ``The ubiquitous XP commutator'' Eur. J. Phys.
\textbf{27}, 225-230, (2006).

\bibitem{dlt1}
A. C. de la Torre, ``On Randomness in Quantum Mechanics'' Eur. J.
Phys. \textbf{29}, 567-575, (2008).

\bibitem{pau} W. Pauli, ``Die allgemeine Prinzipien de Wellenmechanik'',
Handb. Phys. \textbf{24} (2), 83-272 (1933).

\bibitem{wei1}
S. Weigert, ``Pauli problem for a spin of arbitrary length: A simple
method to determine its wave function'' Phys. Rev. A {\bf 45},
7688-7696 (1992).

\bibitem{wei2}
S. Weigert, ``How to determine a quantum state by measurements: The
Pauli problem for a particle with arbitrary potential'' Phys. Rev. A
{\bf 53}, 2078-2083 (1996).

\bibitem{qinf}M. Keyl,
``Fundamentals of quantum information theory'' Phys. Rep. A
\textbf{369}, 431-548 (2002).

\bibitem{unbiaBas1}  I. D. Ivanovic, ``Geometrical description of quantum
state determination'' J. Phys. A, 14, 3241-3245 (1981).

\bibitem{unbiaBas2}W.K. Wootters, and B.D. Fields, ``Optimal state-determination by mutually
unbiased measurements'' Ann. Phys. 191, 363-381 (1989).

\bibitem{dlt4}
D. M. Goyeneche and A. C. de la Torre, ``State determination: An
iterative algorithm'' Phys. Rev. A \textbf{77}, 042116 (2008).

\end{thebibliography}
\end{document}